\documentclass[runningheads]{svmult}

\usepackage{makeidx}   
\usepackage{graphicx}  
\usepackage{subeqnar}  
\usepackage{multicol}  
\usepackage{physprbb}  
\makeindex             

\begin{document}
\title*{The Halo Lithium Plateau: Outstanding Issues}
\toctitle{The Halo Lithium Plateau: Outstanding Issues}
%
%
\titlerunning{The Galactic Halo Lithium Plateau}
%
\author{Sean G. Ryan\inst{1}
\and Lisa Elliott\inst{2}}
\authorrunning{Ryan \& Elliott}
%
%
\institute{Centre for Earth, Planetary, Space and Astronomical Research, Department of Physics and Astronomy,
The Open University, Walton Hall, Milton Keynes, MK7 6AA, United Kingdom s.g.ryan@open.ac.uk
\and
Department of Mathematics and Statistics, School of Mathematical Sciences, Monash University, Clayton, Vic. 3800, Australia
Lisa.Elliott@sci.monash.edu.au}

\maketitle              

\begin{abstract}
We examine outstanding issues in the analysis and interpretation of the halo 
Li plateau.
We show that the majority 
of very Li-poor halo Li-plateau stars (5 out of 8) 
have high projected rotation velocities $v$sin$i$ between 4.7 and 10.4~km~s$^{-1}$. Such stars have very different evolutionary histories to Li-normal plateau stars, and hence cannot be included in studies of Li depletion by normal halo dwarfs. Uncertainties in the effective temperature scale for metal-poor stars continue to challenge the analysis of Li.
\end{abstract}

\section{Introduction}

From observations of 11 main-sequence stars belonging to the Galactic halo, Spite \& Spite 
\cite{ryan:ss82}
concluded that the lithium abundance was essentially independent of metallicity for halo stars hotter than 5600~K, and inferred that the Li abundance 
was ``hardly altered'' from the Big Bang.
Two decades of 
work has followed, increasing the number of stars observed and the range of metallicity that they span, in an effort to establish the primordial Li abundance more securely.

The primordial Li abundance was sought primarily because of its ability to constrain the baryon to photon ratio in the Universe, or equivalently the baryon contribution to the critical density.
In this way, Li was able to complement estimates from $^4$He, the primordial abundance of which varied only slightly with baryon density. Li also made up for the fact that the other primordial isotopes, $^2$H (i.e. D) and  $^3$He,
were at that time difficult to observe and/or interpret. During the late 1990's, however, measurements of D in damped Lyman alpha systems (high column-density gas believed to be related to galaxy discs) provided more reliable constraints on the baryon density than Li could do (e.g. 
\cite{ryan:otksplw01}).
Even more recently, the baryon density has been inferred 
from the angular power spectrum of the cosmic microwave background radiation,  for example from the WMAP measurements 
\cite{ryan:s03}.
We consider the role of Li plateau observations post WMAP.

\section{Interpretation of the Li Observations}

The difficulties in inferring the primordial Li abundance from halo star observations can be separated into two broad categories:
\begin{itemize}
\item establishing the current abundance of Li in the stars being analysed, and
\item determining the degree to which the current Li abundance of a star deviates from the primordial value due to the production of Li in the interstellar gas from which the star formed, or the destruction of Li in the star subsequent to its formation. 
\end{itemize}

The first of these difficulties arises primarily because of uncertainty in the temperature structure of the stars whose spectra are observed, measured, and 
interpreted to obtain Li abundance estimates. The problem 
can in turn be divided into three aspects:
\begin{itemize}
\item the uncertain effective temperature scale of metal-poor stars
\newline Differences as large as 150 to 200~K are frequently noted between the effective temperature scales adopted by different spectroscopists, and such differences can easily include dependences on metallicity and colour  
(e.g. see calibrations considered by 
\cite{ryan:rnb99}, and the final section of the present paper).
A change of 100~K corresponds to 0.08~dex in abundance.
\item the uncertain temperature gradient in the outer layers of the atmosphere
\newline Even if two models have
the same effective temperature, they may 
possess different temperature gradients. Since the 
Li absorption line forms at slightly shallower depths than the range over which the continuum forms, a difference in the temperature gradients in two models means that gas of different temperatures is involved in line formation in each case, resulting in different line strengths. Differences as large as 0.1-0.15 dex have been found from model dependent differences (e.g. 
\cite{ryan:rbdt96}.)
\item the inappropriateness of 1D model atmospheres
\newline A 1D model atmosphere seeks to represent the variation in the 
key physical quantities as a function of distance only along a single line of 
sight. Recently, 3D model atmospheres have been computed for a handful of 
stars, and allow radiation transfer to be computed along a greater number of 
lines of sight into a dynamic medium. One triumph of such models is that the microturbulent velocity which must be included in 1D radiative transfer computations to account for non-thermal motions is no longer required in the 
dynamic 3D atmosphere. The Li abundances derived using 3D LTE models were lower than those from 1D models 
\cite{ryan:ants99},
though NLTE corrections applied to 3D models come close to reproducing 1D LTE results 
\cite{ryan:acb03}.
These early results point the way to a future of more realistic spectral analysis in which 3D modelling becomes the norm.
\end{itemize}

The second difficulty is to relate the observed abundance to the primordial one
\cite{ryan:bs85}.
The question in 1985 was whether the primordial value was close to the abundance measured in young Population I stars, implying that Li had been depleted by an order of magnitude in halo stars, or whether Galactic chemical evolution had provided a source of Li that raised the Galactic abundance to the Population I value from a primordial level close to that observed by Spite \& Spite. The evidence
relied on arguments over the likely mass- (effective temperature-) and metallicity-dependence of depletion mechanisms, and whether a large degree of depletion could in the end bring about a relatively uniform plateau abundance in a range of stars. A slight dependence on effective temperature and metallicity has been claimed 
\cite{ryan:rbdt96,ryan:rnb99},
but not without challenge 
\cite{ryan:bm97,ryan:mr04}.
The elimination of the trends would require the stellar temperatures to be revised by more than the estimated uncertainties, but not by {\it much} more than the estimated range of uncertainties if we have been unlucky and all errors are pushed to their extremes. However, the recent analysis of very high S/N spectra 
\cite{ryan:a05}
likewise finds a metallicity dependence. The preliminary results of the VLT ``First Stars'' study 
\cite{ryan:b03}
foreshadow the forthcoming availability of results to emerge from the full analysis of that data set. The existence of slight metallicity and temperature dependences of the plateau Li abundances, if not due to temperature errors
such as those described above,
indicate that some Galactic production and some stellar destruction of Li has occurred. It therefore becomes crucial to understand the extent of both processes. The fact that dwarfs down to [Fe/H] = $-3.7$ have been measured, and that there are few Galactic objects with lower metallicities, mean that very little Galactic production could have occurred by the time the halo stars formed. That is, 
t
he amount by which the halo plateau value exceeds the primordial one is quite small, $0.11^{+0.07}_{-0.09}$~dex 
\cite{ryan:rbofn00}.
Unfortunately, the size of the other term, i.e. the degree to which halo plateau stars may have destroyed some of the Li they formed with, is less well constrained. That term will be examined in the following section.

\section{Li Destruction in Main-Sequence Halo Stars}

Simple, sometimes called ``standard'', stellar evolution models in which the outer parts of the envelope mix according to standard convective prescriptions predict essentially no depletion of Li by plateau stars over their lifetimes 
\cite{ryan:ddk90}.
However, such models are rather unsatisfactory at explaining some important characteristics of Population I stars, including the depletion of Li and other light elements 
\cite{ryan:dbskvk98}.
Models which invoke more complex, but consequently less well constrained, rotationally-induced mixing seem to reproduce Population I observations better and hence offer another mechanism by which halo plateau stars might destroy Li from a higher initial value to arrive at the observed plateau value 
\cite{ryan:pdd92,ryan:pwsn99,ryan:pswn02}.
The test of these models is whether a range of depletion factors has existed (depending on the initial angular momentum of each protostar), and hence whether a range of plateau Li abundances results. The range of abundances in the sample of 
\cite{ryan:rnb99},
once GCE is allowed for, is very small, having a standard deviation $\sigma$ = 0.031 dex, entirely consistent with the observational uncertainties. 
Ryan et al. 
\cite{ryan:rnb99}
argue that the intrinsic spread in plateau Li abundances at a given metallicity is very small, and they rule out rotationally-induced depletion of Li by more than 0.1~dex.  However, the interpretation of the sample depends crucially on whether one is justified in excluding halo stars which have extremely low Li abundances. Eight such stars are known, having Li abundances that are at least 0.5~dex below those of plateau stars. (Only upper limits can be quoted for their Li abundances.) Pinsonneault et al. 
\cite{ryan:pswn02}
argue is contrast that depletion up to 0.2~dex could still be consistent with the existence of some Li depleted stars.

To determine whether the Li-poor stars have to be considered along with the plateau stars, or alternatively whether their evolutionary histories
are not representative of normal plateau stars, we undertook a more detailed investigation of the Li-poor objects. 
Norris et al. 
\cite{ryan:nrbd97}
\cite{ryan:rnb98}
showed that of four such stars examined, two had element compositions indistinguishable from those of Li-preserving plateau stars of the same metallicity. The other two showed tantalising evidence of unusual abundances of neutron-capture 
elements: G186-26 showed an excess of Sr, Y and Ba relative to Fe, while G139-8 showed a very high [Ba/Sr] ratio, though both [Sr/Fe] and [Ba/Fe] were sub-solar. A study of four further members of the class 
\cite{ryan:fer04,ryan:efrg05}
showed no other element anomalies apart from an elevated Na abundance in one. Clearly there is no widely occurring chemical anomaly in this group of stars other than Li, though the presence of high [Ba/Sr] ratios in G139-8 and G186-26 are suggestive of the involvement of more-massive stars in the evolution of the Li-poor stars or the clouds from which they formed. 

The concentration of Li-poor stars around the main sequence turnoff led 
\cite{ryan:rbkr01}
to speculate that Li-poor stars might be related to blue stragglers. Specifically, they proposed that the same mass transfer processes that produce field blue stragglers should also produce sub-turnoff-mass objects which would be indistinguishable from normal stars except for the absence of Li. The discovery that three out of four Li-poor stars had rotation velocities in the range $v$sin$i$ = 5.5 to 7.6~km~s$^{-1}$, while Li-normal stars had undetectable rotation generally below 3~km~s$^{-1}$, led 
\cite{ryan:rgkbk02}
to conclude that angular momentum transfer in the blue-straggler forming process was responsible for these stars having been spun up to higher than normal rotation speeds. Our new study takes in the remaining four known Li-poor stars, and improves the resolution and S/N of the data on two of those already measured. 
It shows that five of the eight Li-poor stars have rotation velocities in excess of 4~km~s$^{-1}$ 
\cite{ryan:er05}.
Four of these five stars are confirmed binaries, which indicates that mass 
transfer did not result in the complete merger of the components, consistent 
with small transferred masses as inferred by  \cite{ryan:rgkbk02}.
The results are given in Table~\ref{ryan:Tab1} along with previous measurements of similar stars. 
Allowing for the possibility that some of these stars may be viewed at low inclination, i.e. almost pole-on, we cannot rule out the possibility that most are rotating, with $\langle v \rangle \sim  6$~km~s$^{-1}$.

\begin{table}
\caption{Projected rotation velocities of Li-poor stars}
\begin{center}
\renewcommand{\arraystretch}{1.4}
\setlength\tabcolsep{5pt}
\begin{tabular}{lcc}
\hline\noalign{\smallskip}
Star			&$v$sin$i$/km~s$^{-1}$	&$v$sin$i$/km~s$^{-1}$	\\
			&\cite{ryan:rgkbk02}		&\cite{ryan:er05} \\
\noalign{\smallskip}
\hline
\noalign{\smallskip}
Wolf 550 = G66-30	& 5.5 $\pm$ 0.6		&5.6 $\pm$ 0.3		\\
G202-65		& 8.3 $\pm$ 0.4		&8.6 $\pm$ 0.2		\\
BD+51$^\circ$1817	& 7.6 $\pm$ 0.3					\\
HD 97916		& 			&10.4 $\pm$ 0.2	\\
G122-69		&			&4.7 $\pm$ 0.5		\\
BD$-$31$^\circ$19466	& $<$2.2 				\\
G139-8			&			& $<$1.9		\\
G186-26		&			& $<$2.5		\\
\hline
\end{tabular}
\end{center}
\label{ryan:Tab1}
\end{table}

We conclude that the Li-poor stars definitely have different evoutionary
histories to Li-normal plateau stars. A mass-transfer mechanism may explain
the origin of these objects, but irrespective of whether this is the correct explanation, the Li-poor objects cannot be included in studies
of Li depletion mechanisms that affect normal single stars. Consequently we conclude that the small (zero?) intrinsic spread in plateau Li abundances inferred by 
\cite{ryan:rnb99}
is representative of normal halo stars, and thus signifies at most a small depletion in Li, $<$~0.1~dex by the models of 
\cite{ryan:pwsn99}.

\section{Li in the Post-WMAP Era}

If the baryon density of the Universe is well constrained from WMAP observations, what purpose do Li observations in plateau stars now serve? 
The apparent lack of agreement between WMAP and the halo star Li observations could be due to imperfect modelling of big bang nucleosynthesis of Li, for example if the $^7$Be(d,p)2$^4$He rate is wrong 
\cite{ryan:cvdaa04}.
Alternatively, if we are to reconcile the observed Li abundance with the value 
currently inferred from big bang nucleosynthesis calculations, then we have a new constraint on the degree to which it has been processed in halo stars. We can use this measurement of the abundance decrement, in concert with precise measurements of the metallicity- and mass- (temperature-) dependence of surviving Li abundances, to learn much about the way stars' outer envelopes are processed during their lifetimes.

The optimistic sentiments of the penultimate paragraph presume, of course, that our abundance analyses are correct, and that we have not erred by a large factor in inferring the present Li abundances of plateau stars. 
An underestimate of 
the more important systematic uncertainties could 
alter the picture.
A very recent paper 
(\cite{ryan:mr04};
their Figure 2) 
has suggested an upward revision of the effective temperatures of
\cite{ryan:rnb99}
i.e. for stars with [Fe/H] $<$ $-2$,
by typically 200 to 400~K. 
The resulting Li abundances are 
marginally consistent with WMAP when overshooting model atmospheres are used. 
However, temperatures of metal-poor stars based on H$\alpha$ line profiles 
appear not 
to be so much higher than those of \cite{ryan:rnb99}, perhaps only $\sim$120~K 
higher \cite{ryan:a05}.
Additionally, analysing the sample of
\cite{ryan:rnb99},
\cite{ryan:aranb04}
derived spectroscopic effective temperatures on average 80 K {\it cooler} than 
those of 
\cite{ryan:rnb99},
though they were concerned that the excitation temperatures were not well constrained because of the 
small number of Fe~I lines and the low range of excitation potentials 
represented  in the data. WMAP may have reported, but the interpretation of the
Li spectra demands continued effort.

\section{Acknowledgments}
SGR is pleased to acknowledge fruitful discussions and collaborative studies with T. C. Beers, C. P. Deliyannis, A. Ford, J. E. Norris, and M. H. Pinsonneault on issues related to Li, many of which are reflected in this paper.

%

\end{document}